\documentclass{article}
\usepackage{spconf,amsmath,graphicx,hyperref}
\usepackage{booktabs}
\usepackage{comment}
\usepackage{subcaption}
\usepackage{graphicx}
\usepackage{multirow}
\usepackage{amssymb}

\usepackage[table]{xcolor}
\definecolor{mintbg}{rgb}{.63,.79,.95}
\colorlet{lightmintbg}{mintbg!30}
\definecolor{mintbg1}{rgb}{1.0,0.85,0.88}
\colorlet{lightmintbg1}{mintbg1!30}
\usepackage{adjustbox}

\title{Windowed SummaryMixing: An Efficient Fine-Tuning of Self-Supervised Learning Models for Low-resource Speech Recognition}
\name{Aditya Srinivas Menon$^{\star}$ \thanks{$^{\star}$ These authors contributed equally.} \qquad Kumud Tripathi$^{\star}$ \qquad Raj Gohil \qquad Pankaj Wasnik}
  
\address{Media Analysis Group, Sony Research India \\ Email:\{aditya.menon,kumud.tripathi,raj.gohil,pankaj.wasnik\}@sony.com\vspace{-0.2cm}}
\begin{document}
%
\maketitle
\begin{abstract}
Self-supervised learning (SSL) has advanced speech processing but suffers from quadratic complexity due to self-attention. To address this, SummaryMixing (SM) has been proposed as a linear-time alternative that summarizes entire utterances using mean pooling but lacks sufficient local context. In this work, we introduce Windowed SummaryMixing (WSM), which enhances SM by integrating local neighborhood summaries alongside the global summary, maintaining efficiency while improving temporal dependencies. 
Additionally, we introduce a selective fine-tuning approach, replacing self-attention layers in SSL models with WSM blocks and fine-tuning only these blocks in low-resource settings. Our approach improves ASR performance while reducing peak VRAM usage by 40\% in the SSL models. 
WSM blocks have linear-time complexity with enhanced context awareness. Selectively replacing some attention layers reduces compute, memory, and latency, making it ideal for low-resource speech recognition.
\end{abstract}
\section{Introduction}
\label{sec:intro}
Self-supervised learning (SSL) has become a key approach in speech processing, enabling models to extract rich representations from large amounts of unlabeled audio. Models such as wav2vec 2.0 \cite{baevski2020wav2vec} and XLS-R \cite{babu2021xls} achieve state-of-the-art results in Automatic Speech Recognition (ASR), Emotion Recognition, Speaker Verification, Spoken Language Understanding, and Speech Translation \cite{mohamed2022self,nguyen2020investigating}. Their effectiveness lies in generalizing across diverse tasks with minimal labeled data, while self-attention mechanisms allow these models to capture long-range temporal dependencies essential for high-quality speech understanding \cite{NIPS2017_3f5ee243}.

While SSL models excel in performance, their training is computationally demanding due to attention’s quadratic complexity \cite{parcollet2023efficiency}. Linear-time alternatives address this issue, notably SummaryMixing (SM) \cite{summarymixing}, which replaces pairwise attention with a global mean summary combined with current frame features. SM has shown competitive or superior ASR accuracy compared to other linear approaches (Fastformer~\cite{wu2021fastformer}, HyperMixer~\cite{mai-etal-2023-hypermixer}, ContextNet~\cite{han2020contextnet}) with significantly lower compute and memory needs, making it our baseline. However, SM’s global summary lacks adequate local context, limiting fine-grained temporal modeling essential for effective speech representation \cite{han2022improving}.

This work proposes Windowed SummaryMixing (WSM), an improved alternative to SM that enhances efficiency while strengthening speech modeling. WSM augments SM’s global summary with local neighborhood summaries, allowing the model to capture finer-grained temporal dependencies without sacrificing linear-time complexity. To further improve training efficiency, we propose a selective fine-tuning (FT) strategy for SSL models. Instead of updating all model parameters, we replace selected self-attention layers with WSM blocks and fine-tune only these new layers, greatly reducing computational demands.
Fine-tuning an entire SSL model on limited data often leads to overfitting and weak generalization \cite{song2024fd}. Our selective approach mitigates this by focusing updates on the newly introduced WSM layers, cutting memory usage and training time while maintaining or improving accuracy in low-resource settings.
We evaluate the method on diverse datasets, including the Kathbath dataset \cite{kathbath} (Hindi and Tamil), Santa Barbara Corpus of Spoken American English (SBCSAE)  \cite{du2000santa} and Common Voice 7.0 subsets (Mexican Spanish, Mandarin, and Arabic) \cite{ardila2019common}, using SSL models such as wav2vec 2.0 ~\cite{baevski2022data2vec}, HuBERT ~\cite{hsu2021hubert}, data2vec ~\cite{baevski2020wav2vec}, XLS-R \cite{babu2021xls}, mHuBERT \cite{boito2024mhubert}, and MMS \cite{pratap2020massively}. Results show competitive or superior performance with markedly lower computational cost. Our key contributions: (1) Extending SM with local neighborhood summaries for better temporal context at linear complexity; (2) demonstrating that linear-time SSL fine-tuning can match baseline results on low-resource speech recognition.

\section{Related Work on efficient Fine-tuning}
 Previous works have addressed the inefficiency of fine-tuning SSL models from different angles.
 In~\cite{yang2024large}, authors have proposed a framework for evaluating SSL models on downstream tasks. They also compare and study the benefits of a learnable weighted sum layer over using the output of the last layer in downstream tasks.
\cite{zaiem2023fine} investigates techniques to reduce SSL encoder computations during ASR fine-tuning, showing that simple input downsampling reduces computation with minimal performance loss while highlighting the impact of dataset size on robustness. \cite{huang2024voicetuner} introduces a fine-tuning technique by adding additional parameters and retaining the attention module. \cite{shon2024generative} introduced context-aware fine-tuning to improve the self-supervised pre-trained transformer model. Several distillation-based~\cite{wang2023exploring} approaches have emerged that reduce VRAM usage and improve inference time, but these do so at the cost of  performance.
Our proposed Windowed SummaryMixing-based approach addresses the inefficiency problem for SSL fine-tuning, and it is compatible with existing methods \cite{baevski2020wav2vec,hsu2021hubert}.

\begin{figure}
  \begin{subfigure}{0.38\columnwidth}
  \includegraphics[width=\textwidth]{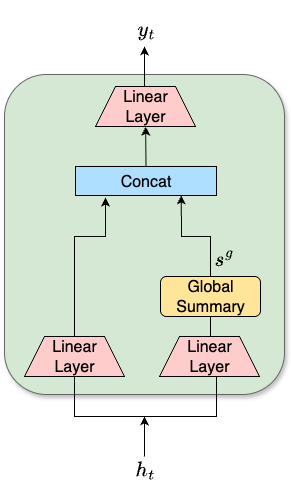}
  \caption{SummaryMixing}
  \end{subfigure}
  \hfill
  \begin{subfigure}{0.55\columnwidth}
  \includegraphics[width=\textwidth]{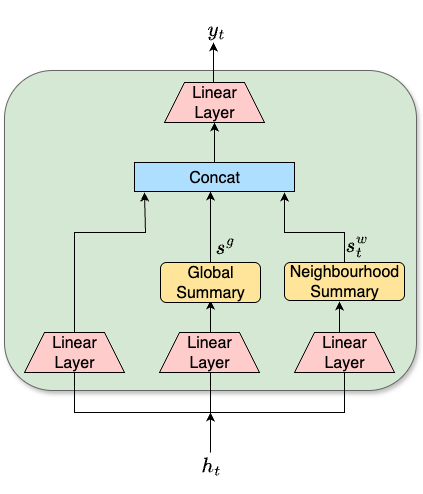}
  \caption{Windowed SummaryMixing} 
  \end{subfigure} 
  \caption{Comparison of the SummaryMixing block (a) and the
 proposed Windowed SummaryMixing block (b). \label{fig1}}
\end{figure}

\section{Windowed SummaryMixing for Fine-tuning SSL}
\subsection{Windowed SummaryMixing}
SummaryMixing~\cite{summarymixing} is a linear-complexity alternative to self-attention. In Figure \ref{fig1}(a, b), the input sequence  of vectors of dimension $d$ with $T$ time steps is represented as $H \in \mathbb{R}^{T \times d}= \{ h_1, h_2, \cdots h_T \}$ and the transformed sequence of hidden vectors are represented as $Y \in \mathbb{R}^{T \times d}= \{ y_1, y_2, \cdots y_T \}$. In Figure \ref{fig1}(a) for the SummaryMixing process, the global summary $s^g$ and transformed output $y_t$ is computed as follows:
\begin{align}
    s^g &= \frac{1}{T} \sum_{t=1}^{T} \texttt{FF}(h_t) \label{eq:summary} \\ 
    {y}_t &= \texttt{FF}(\texttt{concat}(\texttt{FF}(h_t), s^g))
    \label{eq:mixing}
\end{align}
where $\texttt{FF}$ are simple feed-forward layers, and $\texttt{concat}$ stands for concatenation. For more details, refer to the original paper \cite{summarymixing}.
The summary $s^g$ generated in Equation \ref{eq:summary}, uses the mean computed across the whole utterance and is too coarse. Thus in this work, we introduce an additional window-based summary, which generates a neighborhood summary $s^w_t$ in addition to the global summary $s^g$ we already have. 
Specifically, for each frame $h_t$, a mean is computed over the $k$ frames preceding and following the current frame (resulting in a $2*k+1$-sized window), which is then introduced as a third input to the final feed-forward layer. 
We sweep \(k \in \{3,5,7,9\}\) in both mono- and multilingual setups; \(k=5\) consistently yields the best WER, so we fix \(k=5\).
Formally, the operations are defined as follows:
\begin{align}
    s^w_t &= \frac{1}{2k+1} \sum_{j=t-k}^{t+k} \texttt{FF}(h_j) \\
    {y}_t &= \texttt{FF}(\texttt{concat}(\texttt{FF}(h_t), s^g, s^w_t))
\end{align}
It is important to note that, although $s^g$ was common across all the frames, the window-based summary differs from $s^g$ and hence we must use index $t$. For the implementation, we  utilize the SummaryMixing as mentioned in~\cite{summarymixing}.


\subsection{SSL Fine-tuning with Windowed SummaryMixing} \label{sub2.2}
We use wav2vec 2.0, HuBERT, and data2vec as the monolingual and XLS-R, mHuBERT and MMS as the multilingual SSL models in this paper,
since they are a well-established model and popular for SSL research \cite{zaiem2023speech}. Since the listed SSL models are implemented in widely used toolkits such as SpeechBrain \cite{ravanelli2021speechbrain}, results are straightforward
to reproduce. Compared to more recently proposed SSL models,
these SSL models are still competitive for downstream speech processing tasks. 
To enhance the efficiency of SSL models for speech recognition, we replace the traditional self-attention encoder with WSM blocks, which offer a linear-time alternative to attention. While maintaining other model components and training objectives consistent with the original SSL framework, we systematically experimented with different configurations to determine the optimal integration of WSM.

\begin{table*}[t]
  \caption{Speech recognition results (WER in \%) on SSL models with variant of attention encoder block on the Kathbath Hindi and Mexican Spanish datasets. \textit{\textbf{Bold} represents the best result.}}
  \centering
\begin{adjustbox}{width=0.8\textwidth}
  \begin{tabular}{lccccccccccc}
    \hline
    \multirow{2}{*}{\textbf{SSL Model}} &
    \multirow{2}{*}{\textbf{\footnotesize Variant}} &
    \multicolumn{5}{c}{\cellcolor{lightmintbg}\textbf{\footnotesize Kathbath Hindi (WER \( \downarrow \))}} &
    \multicolumn{5}{c}{\cellcolor{lightmintbg1}\textbf{\footnotesize Mexican Spanish (WER \( \downarrow \))}} \\
    & & \footnotesize Last 1 & \footnotesize Last 2 & \footnotesize Last 3 &
    \footnotesize Last 4 & \footnotesize All &
    \footnotesize Last 1 & \footnotesize Last 2 & \footnotesize Last 3 &
    \footnotesize Last 4 & \footnotesize All \\
    \hline
     \multirow{4}{*}{\small{wav2vec 2.0}} & \footnotesize{ SM} & \footnotesize{18.20} & \footnotesize{\textbf{18.18}}& \footnotesize{18.29}& \footnotesize{18.26} & - & \footnotesize{\textbf{34.54}} & \footnotesize{{34.65}}& \footnotesize{35.44}& \footnotesize{34.79} & -\\

          & \footnotesize{ WSM } & \footnotesize{17.89} & \footnotesize{\cellcolor{lightmintbg}\textbf{17.03}}& \footnotesize{18.24}& \footnotesize{18.11} & - & \footnotesize{35.24} & \footnotesize{\cellcolor{lightmintbg1}\textbf{33.97}}& \footnotesize{34.82}& \footnotesize{34.37} & -\\

          & \footnotesize{ Att-PT } & \footnotesize{18.01} & \footnotesize{\textbf{17.95}}& \footnotesize{18.27}& \footnotesize{18.27} & \footnotesize29.82 & \footnotesize{35.52} & \footnotesize{\textbf{34.96}}& \footnotesize{35.30}& \footnotesize{35.02} & \footnotesize 54.44\\

          & \footnotesize{ Att-scratch } & \footnotesize{18.16} & \footnotesize{18.09}& \footnotesize{18.11}& \footnotesize{\textbf{17.90}} & - & \footnotesize{34.99} & \footnotesize{{35.10}}& \footnotesize{\textbf{34.71}}& \footnotesize{35.49} & -\\
        \hline
    \multirow{4}{*}{\small XLS-R} 
    & \footnotesize SM
      & \footnotesize 14.07 & \footnotesize \textbf{13.96}
      & \footnotesize 14.18 & \footnotesize 14.12 & \footnotesize --
      & \footnotesize \textbf{26.38} & \footnotesize 27.52
      & \footnotesize 27.98 & \footnotesize 27.10 & \footnotesize -- \\
    & \footnotesize WSM
      & \footnotesize 13.55 & \footnotesize \cellcolor{lightmintbg}\textbf{13.36}
      & \footnotesize 13.86 & \footnotesize 13.80 & \footnotesize --
      & \footnotesize 27.57 & \footnotesize \cellcolor{lightmintbg1}\textbf{26.42}
      & \footnotesize 27.15 & \footnotesize 26.70 & \footnotesize -- \\
    & \footnotesize Att-PT
      & \footnotesize 13.66 & \footnotesize \textbf{13.60}
      & \footnotesize 13.92 & \footnotesize 13.92 & \footnotesize 22.60
      & \footnotesize 28.44 & \footnotesize \textbf{27.88}
      & \footnotesize 28.22 & \footnotesize 27.94 & \footnotesize 43.46 \\
    & \footnotesize Att-scratch
      & \footnotesize 13.81 & \footnotesize 13.74
      & \footnotesize 13.76 & \footnotesize \textbf{13.55} & \footnotesize --
      & \footnotesize 27.24 & \footnotesize 28.06
      & \footnotesize \textbf{26.92} & \footnotesize 28.69 & \footnotesize -- \\
    \hline
  \end{tabular}
  \end{adjustbox}
  \label{table1}
\end{table*}

\section{Experiments}\label{sec3}
In this section, we describe the baseline architecture, datasets, evaluation metrics, and SSL fine-tuning setup.

\textbf{{Baseline Architecture:}}
\label{sec:baseline}
The SSL models used are based on the transformer encoder architecture. 
Prior work \cite{yang2024large} has shown that using a learnable weighted sum over frozen layers of an SSL model outperforms the conventional evaluation protocol. Following this approach, we adopt the same protocol to train SSL models for low-resource and out-of-domain ASR. In this setting, a lightweight prediction head is mounted on the frozen SSL model, using weighted representations from all frozen layers. These weights are learned jointly during training.  Consequently, we train the weighted layer and lightweight prediction head using SpeechBrain and report these results as our baseline.

\textbf{{Implementation Details:}}
The transformer-based encoder in SSL models consists of multiple hidden layers and attention heads, varying by architecture. To optimize efficiency, we take the baseline SSL architecture with the weighted layer and lightweight prediction head (LSTM) and replace the final two attention layers with linear-time alternatives (see Section \ref{sec4}). Each of these incorporates global and local projection feed-forward layers. A GeLU activation and dropout of 0.1 are applied, while other components align with the model architectures. Fine-tuning is conducted on a single H100 GPU, with a batch size of 16 and 25 training epochs per dataset. Learning rates are set at 1e-3 for the weighted layer and LSTM head, and 3e-3 for unfrozen SM and WSM layers. All systems use character-level tokenization, CTC loss, and no language model.

\textbf{{Datasets and Evaluation Metric:}}
To assess the effectiveness of ASR in low-resource and out-of-domain scenarios which is a key application of SSL models, we utilize datasets from two Indian and four non-Indian languages.
For Indian languages, we select Hindi (hi) from the Indo-Aryan family and Tamil (ta) from the Dravidian family, two of the most widely spoken languages in India. 
These Indian datasets were sourced from the publicly available Kathbath speech corpus.
For non-Indian languages, as discussed in~\cite{yang2024large} in the context of out-of-domain ASR, we utilize Mexican Spanish (es), Mandarin (zh), and Arabic (ar) from Common Voice 7.0, in addition to the SBCSAE, which consists of 60 conversations on diverse topics. The standard splits provided by each corpus are used. 
The performance of the model was evaluated using word error rate (WER), except for Mandarin, where character error rate (CER) is used.


\textbf{Fine-tuning details:}
In the introduced efficient fine-tuning of SSL models, we evaluated four variants to the original attention block: (1) baseline SummaryMixing (SM), (2) proposed Windowed SummaryMixing (WSM), (3) pretrained attention block (Att-PT), and (4) pretrained attention block reinitialized with random weights named as attention scratch (Att-scratch).
For each variant, we investigated different replacement strategies, starting with substituting only the last layer, then extending to the last two layers, and progressively increasing replacements until all self-attention layers were replaced. Each SSL model was trained separately on each dataset with a specific variant, resulting in distinct models for each combination of variant, dataset, and SSL architecture.

\section{Results and discussion} \label{sec4}
This section examines different attention encoder variants for efficient fine-tuning of SSL models on low-resource datasets. 

\subsection{\textbf{Determining Optimal Layer Replacement for Efficient Fine-Tuning}}
Table \ref{table1} compares attention‐replacement strategies for wav2vec 2.0 and XLS-R (Large variants with 24 encoder layers) on the Kathbath Hindi and Mexican Spanish datasets to identify the optimal number of layers for fine-tuning.
We selected one monolingual and one multilingual SSL model to first establish an efficient fine-tuning approach before scaling to all models.
To maintain computational efficiency while representing diverse linguistic features, the evaluation focuses on one Indian and one non-Indian dataset.
In Table \ref{table1}, the column “All” displays results for fine-tuning all layers of the Att-PT variant. This is because fine-tuning an SSL model on low-resource datasets with other variants would reinitialize the weights for each layer, preventing the model from leveraging the pre-trained knowledge of SSL models effectively.


\begin{table*}[t]
  \caption{Speech recognition results (in \%) on finetuning the last two  layers of the SSL models with three variants of attention-based encoder  on the different datasets. \textit{\textbf{Bold} represents the best result.}}
  \centering
\begin{adjustbox}{width=0.8\textwidth}
  \begin{tabular}{lccccccc}
    \toprule
    \textbf{SSL Model} & \textbf{Version} &
    \textbf{SBCSAE (WER \( \downarrow \))} &
    \textbf{hi (WER \( \downarrow \))} &
    \textbf{ta (WER \( \downarrow \))} &
    \textbf{es (WER \( \downarrow \))} &
    \textbf{zh (CER \( \downarrow \))} &
    \textbf{ar (WER \( \downarrow \))} \\
    \hline
    \rowcolor{lightmintbg}\multicolumn{8}{c}{\textbf{Monolingual SSL Models large variants}} \\
    \hline
    \multirow{2}{*}{wav2vec 2.0}
      & Baseline & \textbf{69.24} & 17.31 & 34.66 & 34.30 & 24.10 & 53.82 \\
      & Ours     & 69.26 & \textbf{17.03} & \textbf{32.09} & \textbf{33.97} &
                   \textbf{22.85} & 51.46 \\
    \midrule
    \multirow{2}{*}{HuBERT}
      & Baseline & 70.70 & 18.78 & 36.20 & 29.99 & 22.02 & 48.95 \\
      & Ours     & \textbf{68.54} & \textbf{17.33} & 35.23 & \textbf{28.55} &
                   \textbf{21.56} & \textbf{48.63} \\
    \midrule
    \multirow{2}{*}{data2vec}
      & Baseline & 59.62 & 19.96 & 36.98 & 36.14 & 24.43 & 54.80 \\
      & Ours     & \textbf{59.43} & 19.32 & 36.84 & 35.71 & 21.96 & 51.95 \\
    \hline
    \rowcolor{lightmintbg}\multicolumn{8}{c}{\textbf{Multilingual SSL Models large variants}} \\
    \hline
    \multirow{2}{*}{XLS-R}
      & Baseline & 66.43 & 15.42 & 30.22 & 28.09 & 20.01 & 40.34 \\
      & Ours     & \textbf{63.67} & \textbf{13.36} & \textbf{28.86} &
                   \textbf{26.42} & \textbf{19.98} & \textbf{38.21} \\
    \midrule
    \multirow{2}{*}{mHuBERT}
      & Baseline & 68.35 & 17.55 & 31.82 & 27.38 & 19.82 & 54.54 \\
      & Ours     & \textbf{64.65} & \textbf{15.67} & \textbf{30.77} &
                   \textbf{24.58} & \textbf{18.99} & \textbf{52.31} \\
    \midrule
    \multirow{2}{*}{MMS}
      & Baseline & 55.62 & 16.03 & 29.75 & 29.32 & 18.28 & 42.53 \\
      & Ours     & \textbf{54.86} & \textbf{14.97} & \textbf{28.93} &
                   \textbf{27.44} & \textbf{17.82} & \textbf{39.08} \\
    \hline
  \end{tabular}
  \label{tabel2}
  \end{adjustbox}
\end{table*}


The results show that replacing only the last two layers offers the best trade-off between WER and computational cost, with pre-trained attention (Att-PT) outperforming randomly initialized attention (Att-Scratch). The poor performance of the “All” setting reflects overfitting in low-resource conditions, highlighting the value of selective fine-tuning. Among the tested variants, WSM consistently achieves the lowest WER for Hindi and Spanish, but as WER rises when more than two layers are replaced, subsequent experiments focus on fine-tuning only the final two layers across other SSL models.

\begin{figure}
  \includegraphics[width=0.5\textwidth]{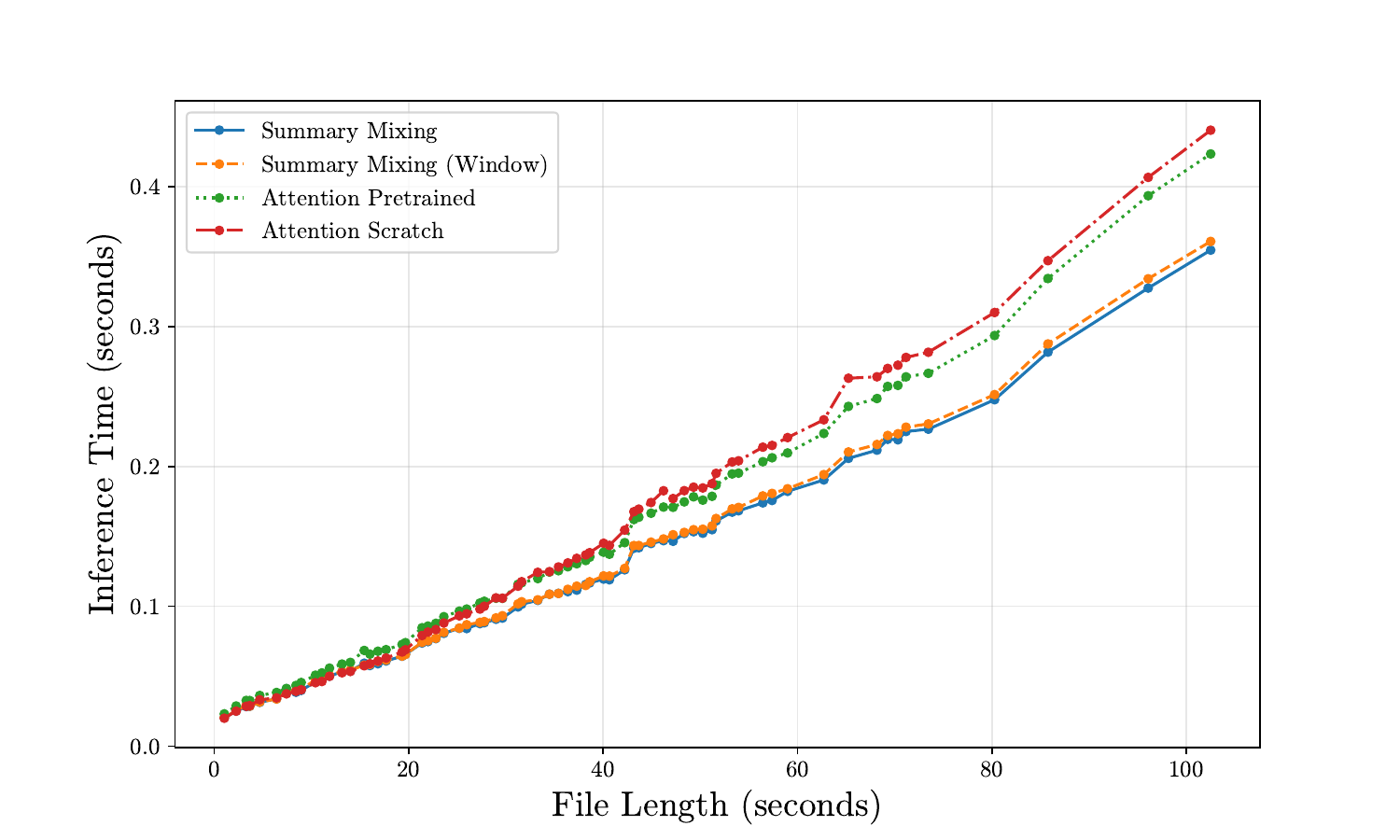}
  \caption{Inference speed of the wav2vec 2.0 with different attention-variants is analyzed as input length varies from 0 to 100 seconds. \label{fig2}}
\end{figure}
\subsection{Speed and Memory Computation} Both SM and WSM variants are significantly more memory-efficient, requiring only 30 GB and 32 GB of GPU VRAM, compared to 50 GB for Att-PT and Att-Scratch. Fine-tuning is also faster with SM and WSM than with Att-PT or Att-Scratch. Thus, SummaryMixing-based methods substantially reduce VRAM usage, by 40\% on multiple runs across different lengths of audio, and training time, making them preferable for resource-constrained environments.

Figure \ref{fig2} compares inference times across multiple attention-variants for varying input lengths. While all variants perform similarly for short sequences (0-10 seconds), WSM-based models become noticeably faster beyond 20 seconds. By 100 seconds, they achieve a 25\% speedup, highlighting their efficiency and suitability for real-time applications requiring low-latency processing.

\subsection{WER Comparison with Baseline}
Table~\ref{tabel2} presents results for both monolingual and multilingual SSL models. To evaluate out-of-domain performance, we first tested monolingual models and then extended experiments to multilingual settings.
Across all settings, replacing the last two attention layers with WSM consistently reduces WER/CER. Monolingual models (wav2vec 2.0 Large, HuBERT Large, data2vec Large) achieve modest but consistent improvements across languages, while multilingual models (XLS-R, mHuBERT, MMS) show even greater relative gains; for example, XLS-R reduces WER for Spanish from 28.09\% to 26.42\% and Arabic from 40.34\% to 38.21\%, and mHuBERT improves both English and Spanish results. These findings demonstrate that WSM-based fine-tuning is effective and scalable for both monolingual and multilingual ASR, providing efficiency gains without sacrificing accuracy.

\section{Conclusion}
In this work, we addressed the quadratic complexity of self-attention in SSL-based ASR models by introducing WSM, a linear-time alternative that enhances SM with local neighborhood summaries. This approach effectively maintains efficiency while improving temporal dependencies. Additionally, we implemented a selective fine-tuning strategy, replacing only the last two self-attention layers with WSM blocks and fine-tuning only these layers in low-resource settings.
Our method consistently outperforms baseline models in WER and CER across multiple dataset. By achieving linear complexity with enhanced context awareness, WSM blocks offers a scalable and computationally efficient alternative to traditional self-attention, making it highly effective for real-time and low-resource speech recognition tasks.
Future work will explore extending WSM to investigate its applicability to other speech-processing tasks. 

\vfill\pagebreak




\bibliographystyle{IEEEbib}
\bibliography{camera_ready}

\end{document}